\documentclass[12pt]{iopart}
\usepackage{iopams}  
\usepackage[utf8]{inputenc}
\usepackage[T1]{fontenc}
\usepackage{orcidlink}
\usepackage{amssymb}
\usepackage{amsmath}
\usepackage{graphicx}
\usepackage{bm}
\usepackage{xcolor}
\usepackage{enumitem}
\usepackage{array}
\usepackage{hyperref}
\usepackage{listings}
\usepackage{comment}
\usepackage{setspace}
\usepackage{datetime}
\usepackage{booktabs}

\begin{document}

\title[Phase Diagram and Crossover Phases  of  Topologically Ordered  Graphene Zigzag  Nanoribbons]{Phase Diagram and Crossover Phases  of  Topologically Ordered  Graphene Zigzag  Nanoribbons: Role of Localization Effects}

\author{
Hoang-Anh Le$^{1,2}$\orcidlink{0000-0002-1668-8984}, 
In-Hwan Lee$^{1,2}$\orcidlink{0000-0002-7619-3780},  
Young Heon Kim$^{1}$\orcidlink{0000-0003-0000-539X} 
and S.-R. Eric Yang$^{1,3}$\orcidlink{0000-0003-3377-1859}}
\address{$^1$ Department of Physics, Korea  University, Seoul 02841, Korea}
\address{$^2$ These authors contributed equally to this work.}
\address{$^3$ Corresponding author: eyang812@gmail.com}

\begin{abstract}
We computed the phase diagram of zigzag graphene nanoribbons as a function of on-site repulsion, doping, and disorder strength. 
The topologically ordered phase undergoes topological phase transitions into crossover phases, which are new disordered phases with non-universal topological entanglement entropy that exhibits significant variance. 
We explored the nature of non-local correlations in both the topologically ordered and crossover phases. In the presence of localization effects, strong on-site repulsion and/or doping weaken non-local correlations between the opposite zigzag edges of the topologically ordered phase.
In one of the crossover phases, both $e^-/2$ solitonic fractional charges and spin-charge separation were absent; however, charge-transfer correlations between the zigzag edges were possible. Another crossover phase contains solitonic $e^-/2$ fractional charges but lacks charge transfer correlations. We also observed properties of non-topological, strongly disordered, and strongly repulsive phases. Each phase on the phase diagram exhibits a different zigzag-edge structure.
Additionally, we investigated the tunneling of solitonic fractional charges under an applied voltage between the zigzag edges of undoped topologically ordered zigzag ribbons, and found that it may lead to a zero-bias tunneling anomaly.

\end{abstract}

\newcommand\ket[1]{\left| #1 \right>}
\newcommand\bra[1]{\left< #1\right|}
\newcommand\bracket [1]{\left( #1\right)}
\newcommand\sbracket [1]{\left[ #1\right]}


\section{Introduction}

Is the formation of a fractional charge~\cite{Lei13,Wilczek03,Arovas,Nakamura01,Barto1,GV2019} a necessary and sufficient condition~\cite{Pach} for topologically ordered insulators~\cite{Wen11} such as fractional quantum Hall systems~\cite{NPlaugh} and interacting disordered zigzag graphene nanoribbons~\cite{Yang} (ZGNRs) with anyonic fractional charges? This issue is related to whether the electron localization effects destroy or enhance the topological order~\cite{Kitaev11,Levin11,Haldane191} and quantization of fractional charges. In a Laughlin state on a sphere (no edges present) and under weak disorder, the added electrons fractionalize and create a quasi-degenerate peak within the gap of the tunneling density of states (DOS).
Electron localization~\cite{Altshuler} is expected to stabilize these fractional charges of the quasi-degenerate  midgap states because these localized quasi-degenerate energy states are spatially separated from each other, as explained in~\cite{GV2000} (if fractional charges are delocalized they  overlap and become ill-defined). However, excessive disorder is considered detrimental to topological order.

In this study, we investigate similar issues with  ZGNRs~\cite{Fujita,Brey2006,Lyang,Pisa1,Cai2,Kolmer,Brey}. 
A recent study showed that weak randomness (disorder) in ZGNRs can generate $e^-/2$ solitonic fractional charges~\cite{Heeger,Yang2019,yang1}, which is a disorder effect closely related to the change in the disorder-free symmetry-protected topological insulator of ZGNRs to a topologically ordered~\cite{Yang2020}  Mott-Anderson insulator~\cite{Dob,Belitz,Byczuk}. These systems have a universal value for topological entanglement entropy (TEE)~\cite{Kitaev11,Levin11} in the weak-disorder regime~\cite{Yang2021}. The shape of entanglement spectrum is also found~\cite{Yang2022} to be similar to the DOS of the edge states, as expected of topologically ordered systems~\cite{Haldane191}.
In interacting disordered ZGNRs, the gap is  further filled by edge states ~\cite{Efros,Yang2019} with an increasing strength of the disorder potential. 

\begin{figure}[hbt!]
    \centering
    \includegraphics[scale=0.45]{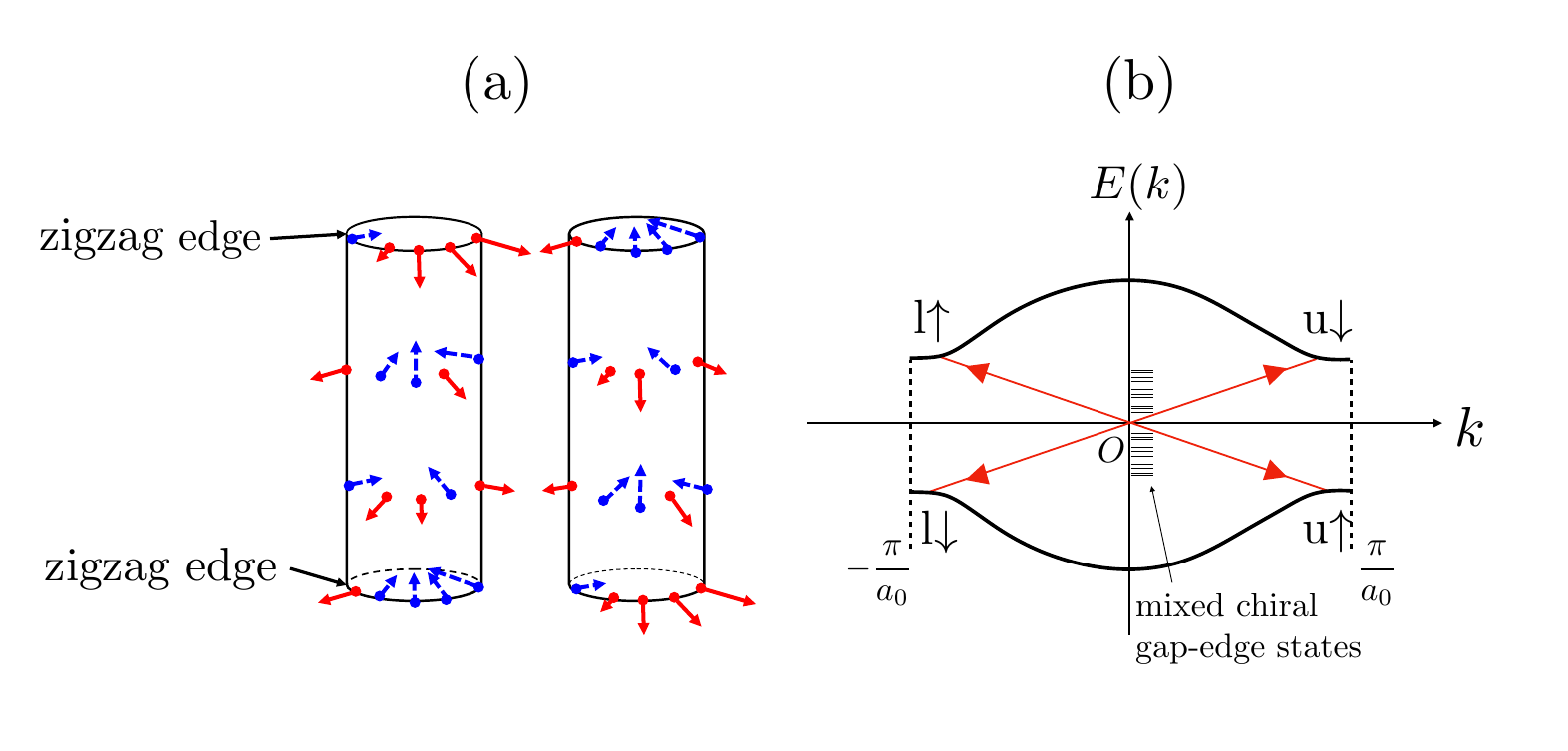}
    \caption{(a) Two degenerate disordered ground states are illustrated on the surface of a cylinder. The site spins of these ground states are reversed. The expectation values of site spins pointing outward (inward) normal to the cylinder surface are represented by solid red (dashed blue) vectors.  Nearly zero-energy  gap-edge states exist on the zigzag edges.
(b) Chiral edge states {are present} near the Brillouin zone boundaries. Here, $\textrm{u} (\textrm{l})$ stands for  a chiral  state localized on the upper  (lower) zigzag edge and $\sigma=\uparrow, \downarrow$ for the spin components along the $z$-axis.  Short-range disorder potential couples two nearly chiral edge states: a $(\textrm{u},\sigma)$ chiral edge state is coupled to a $(\textrm{l},\sigma)$ chiral edge state. This coupling process generates a mixed chiral gap-edge state with two fractional charges.}
    \label{cylinder}
\end{figure}

In the absence of disorder, the ground states have opposite edge site spins, but each edge is ferromagnetically polarized ~\cite{Fujita, Brey2006, Lyang, Pisa1, Cai2, Kolmer, Brey}.
 In the presence of disorder, a spin reconstruction of the zigzag edges can occur~\cite{Yang2020}.  Nonetheless, a topologically ordered ZGNR has two degenerate ground states, see Figure \ref{cylinder}(a).
Mixed chiral gap-edge states (see Figure \ref{cylinder}(b)) play an important role in  this effect.
A short-range disorder potential  couples two nearly chiral gap-edge states residing on opposite zigzag edges~\cite{Yang2019,Lima2012}, and mixed chiral gap-edge states with split probability densities may form 
to display $e^-/2$ fractional semion charges~\cite{Canri}. These states, with midgap energies, are solitonic, with half of the spectral weight originating from the conduction band and the other half from the valence band~\cite{Heeger}. Note that a mixed chiral gap-edge state has a nonzero fractional probability at the A- and B-carbon sites. In other words, it is split into two nonlocal parts, each residing on the edges of the A or B sublattice. The formation of mixed chiral gap-edge states is a nonperturbative instanton effect~\cite{Yang2019}. (They are akin to the bonding and antibonding states of a double quantum well.) 
It should be noted that well-defined $e^-/2$ solitonic fractional charges in the weak-disorder regime are emergent particles, i.e., they have new   qualitative features and appear only in sufficiently long ribbons.
In a weak-disorder regime, the number of fractional charges is  {\it proportional} to the length of zigzag edges. 

Although weak disorder  leads to formation of fractional charges, strong disorder may destroy them. Similar to fractional quantum Hall systems, the topological order of a ZGNR is not  immediately destroyed upon doping because electron localization partially suppresses quantum fluctuations between quasi-degenerate mid-gap states. The system may still be an insulator with a fractional charge.  However, in the presence of strong disorder or doping, zigzag edge antiferromagnetism is expected to diminish, thereby affecting the topological order. 
(In the doped region, a disordered anyon phase with a distorted edge spin density wave was {discovered}~\cite{Yang2022}.) These results suggest the possibility of multiple {\it topological phase transitions} in the zigzag ribbons. What is the nature of these topological phase transitions and the physical properties of the ground states? Does the presence of a fractional charge imply a universal value of the TEE? Does the TEE become nonuniversal and vary~\cite{Huang2011}  with an increase in the disorder strength or doping level?   

The main results of our paper are as follows: the phase diagram of ZGNRs  is found in the parameter space comprising  on-site repulsion ($U$), disorder strength ($\Gamma$), and doping concentration ($\delta N/N_s$) ($\delta N$ and $N_s$ are, respectively, the number of doped electrons and the total number of sites in the  ribbon). 
We identified a number of different phases, including those with topological order, quasi-topological order, and no order.  Each of these phases is defined by the value of TEE $\beta$ and its variance. These properties of $\beta$ are related to the presence or absence of charge fractionalization and charge transfer correlations between  zigzag edges.  When both of these properties are present, along with correlations leading to spin-charge separation, $\beta $ is universal with small variances. In this case, we  investigated the tunneling of fractional charges under an applied voltage between the zigzag edges of undoped topologically ordered zigzag ribbons and  observed a possible zero-bias anomaly. 
The other two types of phases  fall into the category of quasi-topological order. We refer to these phases as crossover phases, in which the variance of $\beta$ is significant.  In one of these phases, both $e^-/2$ fractional charges and spin-charge separation are absent; however, the charge transfer ($\pm e^-/2$) correlations exist between the zigzag edges.
Another phase may contain weakly stable  $e^-/2$ fractional charges but  lacks charge transfer correlations between the zigzag edges.  We also investigated the ground state and zigzag edge properties of the various crossover and nontopological phases.

\section{Model Hamiltonian}

The following mechanisms can all  give rise to fractional charges: the coupling between the valleys mediated by short-range  scatterers~\cite{Yang2019} and the sublattice mixing facilitated by the alternation of the nearest neighbor hopping parameters~\cite{Yang2020}. Here we will consider  the effect of short-range scatterers only. 
The self-consistent Hartree-Fock approximation works well for graphene systems~\cite{Lyang,Pisa1,Stau}. The  Hartree-Fock Hamiltonian of a ZGNR with length $L$  and width $W$ is 
\begin{align}
H_{\textrm{MF}}&=-t\sum_{n.n.,\sigma} c^{\dag}_{i,\sigma}c_{j,\sigma} +\sum_{i,\sigma} V_ic_{i,\sigma}^{\dag}c_{i,\sigma} \nonumber\\
&+U\sum_i[ n_{i,\uparrow}\langle n_{i,\downarrow}\rangle +n_{i,\downarrow}\langle n_{i,\uparrow}\rangle-\langle n_{i,\downarrow}\rangle\langle n_{i,\uparrow}\rangle] 
+ \sum_i [s_{ix} \langle h_{ix}\rangle+s_{iy} \langle h_{iy}\rangle],
\label{MFhspin}
\end{align}
where the site index is given by $i=(k,l)$ ($k$ labels sites along the ribbon direction and $l$ along the width), $c^\dagger_{i, \sigma}$ and $n_{i_,\sigma}$ represent creation and occupation operators at site $i$ with spin $\sigma = \left\{ \uparrow, \downarrow \right\}$, respectively (periodic boundary conditions are used along the ribbon direction). The site spin operators are given by  $s_{i x (y)} = \frac{1}{2}( c^\dagger_{i, \uparrow}, c^\dagger_{i, \downarrow}) \sigma^{x (y)} ( c_{i, \uparrow}, c_{i, \downarrow})^T $, where $\sigma^{x (y)}$ is the conventional Pauli matrix. The first term represents the kinetic energy with hopping parameter $t$,  with $n.n$ denoting the summation over the nearest-neighbor sites. The second term represents the short-range impurities parameterized by $V_i$, which is randomly chosen from the energy interval $\left[ -\Gamma, \Gamma \right]$. Throughout this study, the density of the impure sites is fixed at $10 \%$.  $U$ denotes the on-site repulsive strength. The last term in (\ref{MFhspin}) represents self-consistent ``magnetic fields", where $\langle h_{i x} \rangle = -2 U \langle s_{i x} \rangle$ and $\langle h_{i y} \rangle = -2 U \langle s_{i y} \rangle$. (These fields are present only in doped ZGNRs. In the initial stage of the Hartree-Fock iteration, the values of $\langle h_{i x} \rangle$ and $\langle h_{i y} \rangle$ can be selected from small random numbers). In the presence of these fields, the Hartree-Fock eigenstates are mixed spin states. 
The Hartree-Fock single-particle states $\ket{k}$  ($k=1,2,\ldots,2N_s$) can be written as a linear combination of site states $\ket{i,\sigma}$.  In  the language of second quantization  this is equivalent to 
\begin{equation}
a_k=\sum_{i,\sigma} A_{k,i,\sigma}c_{i,\sigma}.
\label{asumc}
\end{equation}
These magnetic fields are rather small for the  disorder strength and   doping level considered in this study.

There may be several nearly degenerate Hartree-Fock ground states.  We select the Hartree-Fock initial  ground state such that $\langle n_{i,\sigma}\rangle$ represents a  paramagnetic state with a small spin splitting.
In addition, we choose small random { values for} $\langle h_{i x} \rangle$ and $\langle h_{i y} \rangle$ ({these values} do not significantly affect the final results).
The Hartree-Fock matrix dimension scales with  the number of carbon atoms, which is typically  less than $50,000$.  The Hartree-Fock eigenstates and eigenenergies are self-consistently computed,  requiring approximately $20$ iterations. The TEE is computed using the disorder-averaging results from numerous disorder realizations.  Here, we used GPU to speed up the solution of the Hartree-Fock matrix. The GPU calculations were  computationally intensive and were performed on a supercomputer. In the presence of disorder and in the low-doping region, the  obtained Hartree-Fock ground-state properties with solitons are in qualitative agreement with those of the density matrix renormalization group in the matrix product representation~\cite{Yang2022}.   However, quantum fluctuations are not included in the Hartree-Fock calculation, so it may overestimate the stability of fractional charges. (In this work we do not investigate the high doping region. Obtaining density matrix renormalization group results in this region is challenging because the computation is rather time consuming, making it impossible to determine the true ground state among the nearly degenerate Hartree-Fock ground states.)

 Note that the Mott gap in the absence of disorder, denoted by $\Delta$, is well-developed only when the length of the ribbon is much greater than its width $L\gg W$ (for a ribbon with $L\sim W$, its excitation spectrum is similar to that of a gapless two-dimensional graphene sheet~\cite{Neto}). The localization properties of ZGNRs are unusual because both localized and delocalized states can exist~\cite{Yang2019,Lima2012}.  Gap-edge states with energy $|E|\sim \Delta/2$ can have localization lengths of approximately $W$ and have significant overlap with each other.

\section{Effects of Anderson localization}\label{Loc}

 The following simple toy model, although not realistic, is useful in understanding physics of localization~\cite{Altshuler}.  Consider two localized orbital wave functions, $\phi_{1}$ and $\phi_{2}$, each having eigenenergies $\epsilon_1$ and $\epsilon_2$. 
Suppose these two states are coupled by an energy $I$.  The Hamiltonian of the coupled system is 
\begin{equation}
H= \begin{pmatrix}
\epsilon_1   & I   \\
I  & \epsilon_2 
\end{pmatrix}.
 \label{Twotwo}
\end{equation}
For the non-resonant case, where $\epsilon_2 - \epsilon_1 \gg I$, the perturbed eigenstates are given by
\begin{equation}
\psi_{1,2}\approx\phi_{1,2}+ \mathcal{O} \left(\frac{I}{\epsilon_2  -\epsilon_1  }\right)\phi_{2,1}.
\end{equation}
Thus, when the energy difference $\epsilon_2 - \epsilon_1$ is greater than the hopping energy $I$, the states remain nearly unchanged,  i.e., they stay localized.
In the opposite resonant case, where  $\epsilon_2 - \epsilon_1 \ll I$, the states $\phi_{1}$ and $\phi_{2}$ become strongly perturbed and transform into resonant states given by
\begin{equation}
\psi_{1,2}\approx\phi_{1,2}\pm \phi_{2,1}.
\label{Reson}
\end{equation}
Therefore, when the energy difference is sufficiently small, the states may become delocalized,  and the overlap between the localized  states  is significant. Symmetric and antisymmetric states of a quantum double well serve as good examples of these delocalized states.

\begin{figure}[hbt!]
\centering
\includegraphics[width = 0.6 \textwidth]{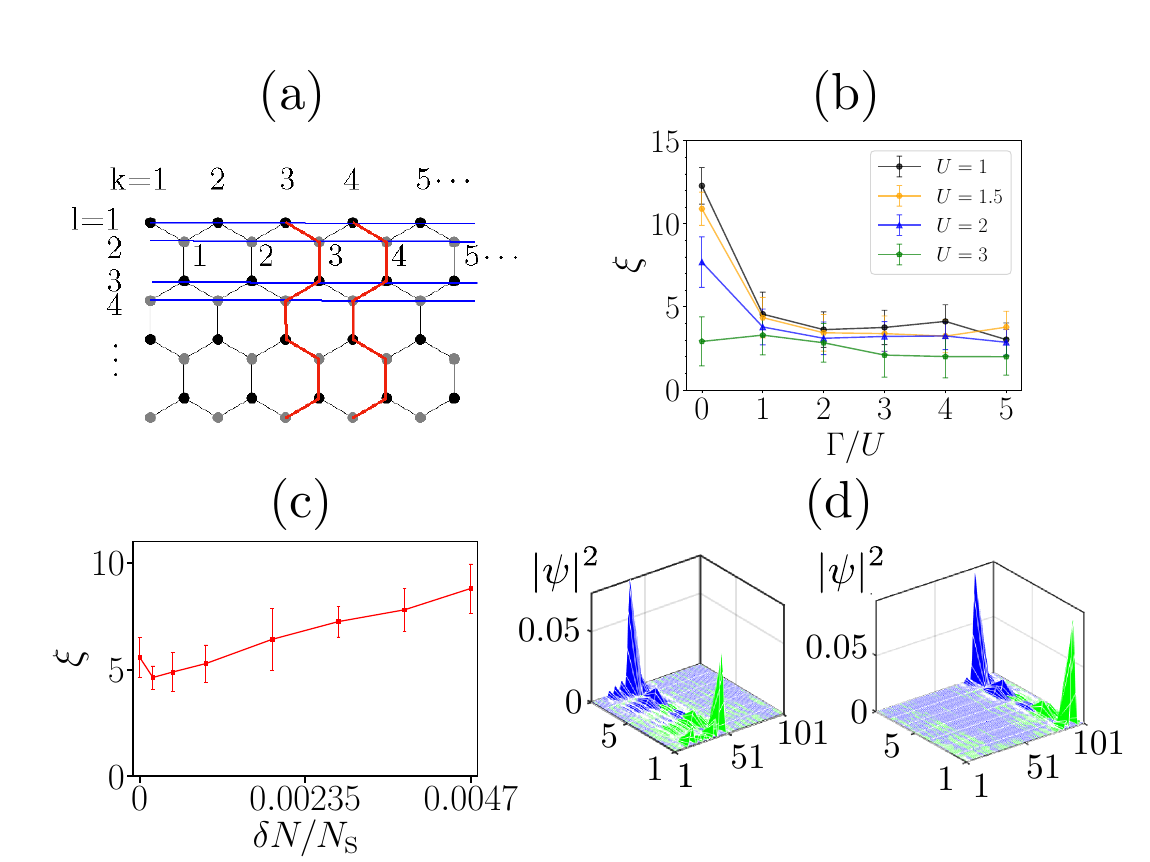}
\caption{(a) Schematic figure of a ZGNR which consists of two triangular sublattices A (black sites) and B (grey sites). Each site of a ZGNR is labeled by two indices $(k,l)$, corresponding to the blue and red carbon lines.  The line $l=1$ comprises zigzag edge sites, which are not connected to each other through the hopping term of the Hamiltonian. (b) Correlation length $\xi$ is estimated by fitting the correlation function with an exponential function along the zigzag edges. In this case, the ribbon length and width are $(L, W) = (100, 64)$. For each data point, we created $N_D = 5$ disorder realizations.
(c) Dependence of correlation length on doping concentration. Each point is calculated with $(U,\Gamma) = (t, t)$, and number of doped electrons are: $\delta N = 0, 2, 5, 10, 20,30 $. At each data point, $N_D = 6$ disorder realizations were created.
(d) Fractional charge probability densities of two quasi-degenerate mixed chiral gap-edge states with $E/(\Delta/2) = -0.005$ (left) and   $E/(\Delta/2) = 0.004$ (right) taken from a single-disorder realization in the weak disorder regime at low doping. These states have the same spin. The Mott gap in this case is $\Delta = 0.2t$. The green (blue) surface refers to the probability density of sites A (B). On-site repulsion, disorder strength,  and number of doped electrons are $(U,\Gamma, \delta N) = (t,0.01t, 3)$. The geometry of the ribbon is $(L,W) = (101,8)$.}
\label{default2}
\label{correlation_length}
\end{figure}

Anderson localization plays a crucial role in the quantization of fractional charges~\cite{GV2000}. The effects of Anderson localization can be described using self-consistent  Hartree-Fock approximation~\cite{Yang1995,Mac1}. The first important effect of Anderson localization  in undoped ribbons in the presence of on-site repulsion is that quasi-degenerate localized  midgap states are spatially {\it separated}~\cite{Altshuler,Efros}, leading to well-defined fractional charges~\cite{GV2000}. (In low-doped ZGNRs, added electrons fractionalize and form a narrow peak in the DOS near $E=0$ consisting of quasi-degenerate localized states ~\cite{Yang2022}.) 
Figure \ref{default2}(d) displays the probability densities of such two midgap states carrying fractional charges.
These gap-edge states are {\it mixed chiral gap-edge states}~\cite{Yang2019}, whose probability densities  peak at the two edges and rapidly decays inside the ribbon.
Note that these states do not overlap with each other. Non-interacting electrons of disordered ZGNRs also display mixed chiral gap-edge states near $E=0$. 
However, in the weak-disorder regime, the overlap between nearly degenerate states is small but not negligible.
Thus well-defined fractional charges do not readily form in non-interacting disordered ZGNRs.

Another important effect is as follows:  it affects the value of the correlation length, which is determined from the entanglement entropy of an area $A$ by computing the reduced density matrix of region $A$.
The Hartree-Fock correlation function~\cite{Yang2021,Peschel119} between $i\in A$ and $j\in A$ decays exponentially as a function of distance $x$ between $i$ and $j$
\begin{equation}
C(x) =  C_{i\uparrow, j\uparrow} =  \langle \Psi \vert c^\dagger_{i\uparrow} c_{j\uparrow} \vert \Psi \rangle  \sim \exp \left(- \frac{\vert x\vert}{\xi} \right),
\end{equation}
where $\Psi$ represents the Hartree-Fock ground state of the ZGNR, and  $\xi$ represents the correlation length.
By inverting the relation given in Eq.(\ref{asumc}), we can express $c_{i\sigma}$ as a linear combination of $a_k$. 
To ensure accurate computation, the  diameter of the area  must be larger than the  correlation length~\cite{Pach,Bal}.

The results of correlation length $\xi$ are shown in Figure \ref{correlation_length}. 
For disordered cases with $\Gamma \neq 0$, the correlation length is obtained by averaging over several disorder realizations. Anderson localization leads to a reduction in the correlation length when compared to disorder-free ribbons, as demonstrated in Figure \ref{correlation_length}(b). (Disorder-free ZGNRs exhibit a large correlation length for small $U$.)  Therefore, compared to the non-disordered case, we can use a smaller Wilson loop~\cite{Levin11}   to calculate the TEE of disordered interacting ZGNRs.   In contrast, doping results in an increased correlation length, as depicted in Figure \ref{correlation_length}(c).   We stress that, in computing the correct TEE, it is important that \textit{the correlation length is small} in comparison to the dimensions of the Wilson loop.

\pagebreak

\section{Phase Diagram }

\begin{figure}[hbt!]
 \centering
\includegraphics[width = 0.9\textwidth]{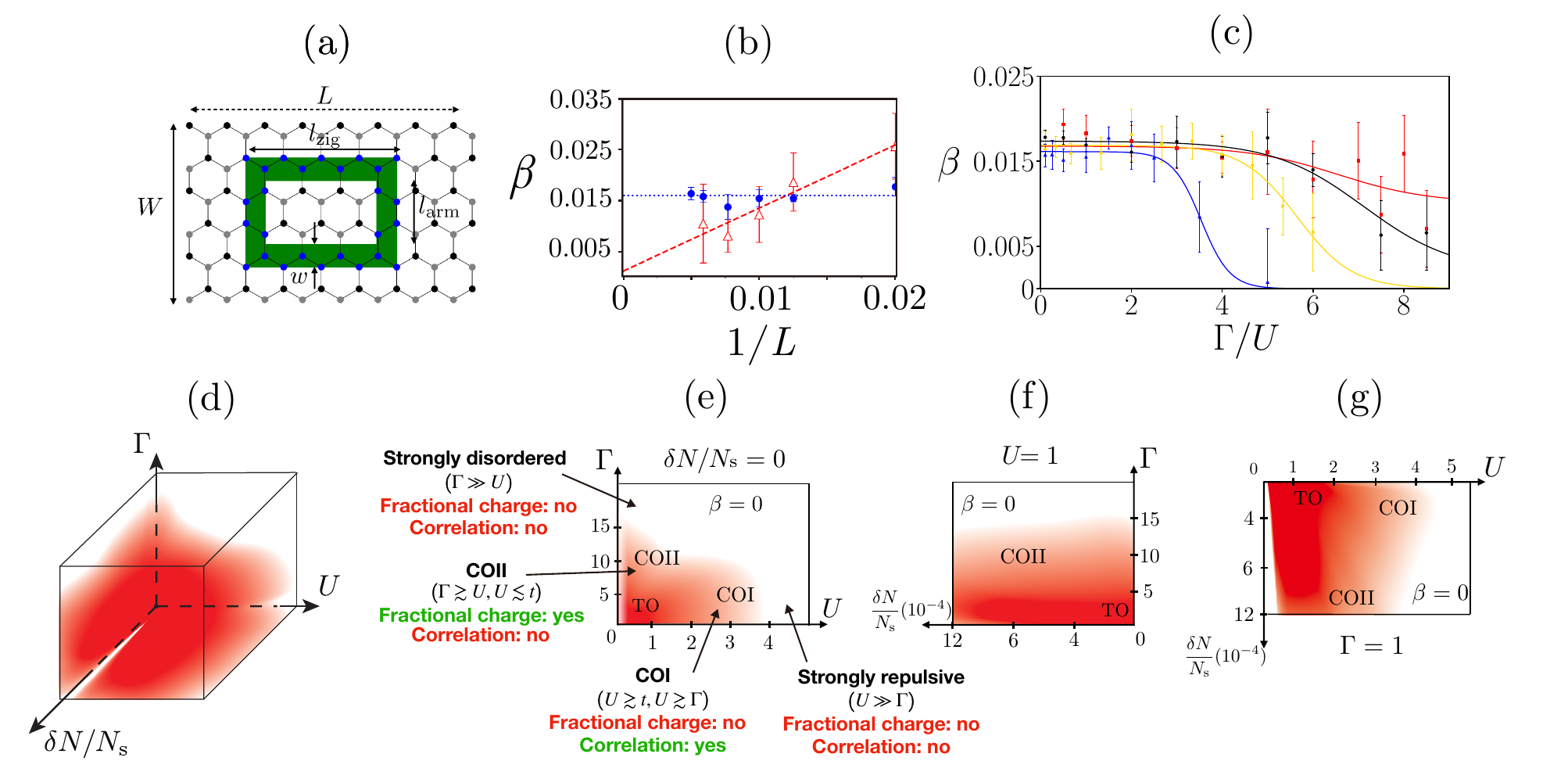}
\caption{(a) A schematic to explain the geometric parameters of a ribbon. $W$ represents the ribbon width, {accounting} both $A$ and $B$ sites, marked with a solid line. $l_{\textrm{zig}}, l_{\textrm{arm}}$, and $w$ are respectively, zigzag length, arm length, and width of the Wilson loop. (b) Finite-size scaling analysis of TEE in universal regime (blue circles, $(U,\Gamma,  \delta N) = (t, t, 0)$) and strongly disordered regime (red triangles, $(U,\Gamma,  \delta N) = (t,15t,  0)$). Ribbon sizes used in the analysis are $(L, W, w, l_{\textrm{zig}}, l_{\textrm{arm}})$: $(200, 172, 12, 171, 68)$,  $(170,144, 10, 145, 60)$, $(130,112, 7, 111, 46)$, $(100,64, 4, 81, 28)$, $(80,68, 4, 69, 24)$; and $(50,32, 2, 41, 12)$. At each data point,  $N_D = 50$ disorder realizations were made.
(c) TEE calculation for an undoped ZGNR $(\delta N = 0)$: red squares correspond to TEE with $U = 0.5t$, black circles are with $U = t$, yellow hexagons are with $U = 1.5t$, and blue triangles are with $U = 2t$. The values of TEE were estimated {through} finite-size scaling. For each data point, the number of disorder realizations is $N_D \sim 30$. The value of TEE is zero at $\Gamma/U = 0$.  
(d) {A schematic} phase diagram of finite-length ZGNRs obtained using the Hubbard model. Regions with degraded color indicate crossover phases.
(e), (f), (g) Projections of phase diagrams onto 2D planes. 
Acronyms TO and COI (COII) stand for topologically ordered and crossover phase I (II), respectively. Here $N_{\textrm{s}} = L \times W$ represents the total numbers of carbon sites of the ribbon, and $\delta N/N_{\textrm{s}}$ represents the doping concentration. The degradation of color represents the variance of TEE (red means small variance), and the white-color regions represent trivial TEE ($\beta = 0$). }
\label{phase_diagram_2}
\label{finitesize}
\end{figure}

Topological order can be detected by investigating the TEE ($\beta$)~\cite{Kitaev11,Levin11} within the Hartree-Fock approximation~\cite{Yang2021}.  We begin by selecting a set of values for $(L, W, w, l_{\textrm{zig}},l_{\textrm{arm}})$, as defined in Figure \ref{finitesize}(a), to compute $\beta$. 
Subsequently, we increment these quantities by the same ratio, and a new $\beta$ is computed. This process is repeated several times (see \cite{Yang2021} for details).
We employ finite-size scaling analysis to extract the TEE's value in the limit as $L$ approaches infinity (see Figure \ref{finitesize}(b)).
We discretize the parameter space $(\Gamma,U,\delta N)$ into a three-dimensional grid, and at each grid point, we compute $\beta$ (see Figure \ref{finitesize}(c)).
The resulting three-dimensional phase diagram is depicted in Figure \ref{phase_diagram_2}(d).
We observe that $\beta$ can have three types of values: (i) A universal value in the topologically ordered phase, (ii) nonuniversal values with large variances in the crossover  phases,  and (iii) a zero value in the normal-disordered phase. Projections of the phase diagram, including the $U$-$\Gamma$,
$\Gamma$-$\delta N$, and $U$–$\delta N$ planes, are presented in Figures \ref{phase_diagram_2}(e)-(g).

In undoped ZGNRs, a  topologically ordered phase is observed in regions $\Gamma/U \lesssim 1$ and $U\lesssim t$, see Figure \ref{phase_diagram_2}(c). The topological phase transition into the  symmetry-protected phase at $\Gamma=0$ is \textit{abrupt}, consistent with findings in \cite{Yang2021} (TEE of the  symmetry-protected phase phase is zero).  There are also other topological phase transitions, but they are smooth transitions with crossover regions.   Note that the system is not topologically ordered at $U=0$: our Hartree-Fock results show that fractional charges overlap with each other and  charge-transfer correlations are absent.
Figures \ref{phase_diagram_2}(e)-(g) reveal the presence of crossover regimes lying beyond the  topologically ordered phase with an increase in the disorder, doping, and interaction strength. 
The phase boundaries between topologically ordered and normal phases are ``blurred,'' indicating the presence of crossover phases (comprising two types, crossover phase I and crossover phase II).  The numerical results of the TEE are shown in  Figure~\ref{phase_diagram_2}(c).
The error bars in this figure include not only random fluctuations caused by disorder but also the uncertainties arising from the extrapolation process of the finite scaling analysis.
As $\Gamma/U$ increases, $\beta$ decreases (indicated by the red line in Figure \ref{phase_diagram_2}(b)). The value of the TEE thus changes across a crossover phase.
Within this phase, $\beta$ exhibits a large variance, but the average values are not zero, which implies that the topological order is not completely destroyed.  In this regime, the TEE becomes nonuniversal and decays.

One can use a different but equivalent 
procedure to determine the phase diagram. We have verified that a similar phase diagram can be obtained by analyzing the presence of fractional charges and nonlocal correlations between the opposite zigzag edges. For each grid point $(\Gamma,U,\delta N)$ in the parameter space, we find the ground state and investigate whether the gap-edge states display fractional charges and whether nonlocal correlations exist between the opposite zigzag edges.  By utilizing this method, we have successfully recovered the phase diagram depicted in Figure \ref{phase_diagram_2}(d).

\section{Topologically ordered phase}

\begin{figure}[hbt!]
\begin{center}
\includegraphics[width = 0.8 \textwidth]{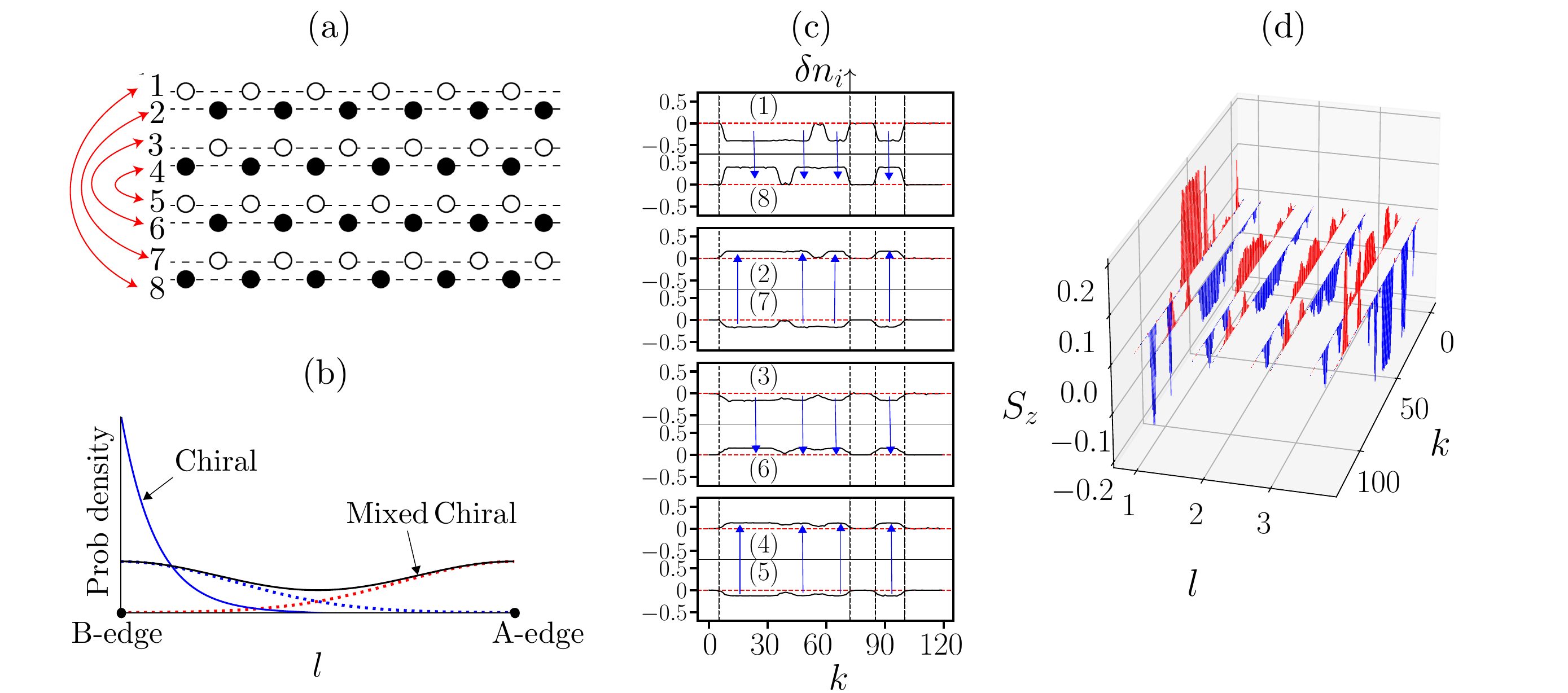}
\caption{(a) A ZGNR  comprises lines of carbon atoms, labeled $1, 2, 3,\ldots, 8$. Red arrows connect two opposite lines that support charge transfers.
(b) Schematic of the probability densities of chiral and mixed chiral gap-edge states. A mixed state (black) comprises both A-(red) and B-(blue) probability components, while a chiral state has only one component. The depicted mixed state in the figure illustrates the possibility of overlapping A- and B-components; however, it is worth noting that they can be also non-overlapping.  
(c) Change in the spin-up occupation number $\delta n_{i,\uparrow}=n_{i,\uparrow}-n^0_{i,\uparrow}$ of a ribbon with dimensions $(L, W) = (120, 8)$ and parameters $(U,\Gamma,  \delta N) = (2t, 0.2t,  0)$ (here $n^0_{i,\uparrow}$ denotes the occupation numbers in the absence of disorder). 
Occupation numbers are computed for a single disorder realization. 
The $x$-axis label $k$ denotes sites along each carbon line $l = 1,2,\ldots,8$ (see Figure \ref{correlation_length}(a)). Blue arrows signify where charge transfer takes place, and black dashed lines indicate where the occupation number changes. Similar results were also found for spin-down occupation numbers.
(d) Site spins computed with parameters similar to (c) are displayed.
}
\label{TO}
\end{center}	
\end{figure}

We elucidate the nature of non-local correlations in topologically ordered ZGNRs. Figure \ref{TO}(a) presents a ZGNR consisting of 8 carbon lines labeled $l = 1, 2, \ldots, 8$.
In each pair of carbons lines $(1, 8)$, $(2, 7)$, $(3, 6)$, and  $(4, 5)$, an increase/decrease in the occupation number of one line is correlated with a decrease/increase in that of the other line (as illustrated by lines 1 and 8 in Figure \ref{TO}(c)).
This correlation is not limited to the zigzag edges but extends to other carbon lines {\it inside} the ribbon, away from the edges.
The corresponding site spins of the ribbon are shown in Figure \ref{TO} (d).
These non-local correlations are attributed to mixed chiral gap-edge states (which can decay relatively {\it slowly} from the zigzag edges, unlike the fractional edge states), as schematically represented in Figure \ref{TO}(b).
Changes in the occupation numbers $\delta n_{i,\uparrow}$ and  $\delta n_{i,\downarrow}$ of an edge often coincide at nearly the same values as $k$, which labels the site position along the ribbon direction. This effect can lead to $n_{i,\uparrow}\approx n_{i,\downarrow}$ of the occupation numbers in the presence of disorder, resulting in $s_i\approx 0$, i.e., the appearance of spin-charge separation around a site on one of the edges~\cite{Yang2020}.

The following points should be noted as well. The results in Figure \ref{phase_diagram_2}(c) show that 
the variance of $\beta$ decreases as we approach the singular limit of $\Gamma/U\rightarrow 0$. (Additional numerical results confirm this conclusion.) This result is consistent with the previous finding that fractional charge of a midgap state  becomes {\it accurate} in the weak disorder regime and in the thermodynamic limit (as discussed in \cite{Yang2022}).
In the opposite limit of $\Gamma/U \gg 1$, the value of the TEE becomes non-universal and decreases with increasing $U$ (as shown in Figure \ref{phase_diagram_2}(c)).  In addition, the functional dependence of DOS on $E$ in the universal region is given by an exponentially suppressed  function, a linear function~\cite{Yang2020}, or something in between. The actual shape of the DOS is determined by the competition between the strength of disorder and the on-site repulsion~\cite{Byczuk}. For instance, the DOS is linear for $(U, \Gamma, \delta N) = (2t, 0.5t,  0),$ but it is exponentially suppressed in the weak disorder limit. 

\begin{figure}[hbt!]
\begin{center}
\includegraphics[width = 0.37 \textwidth]{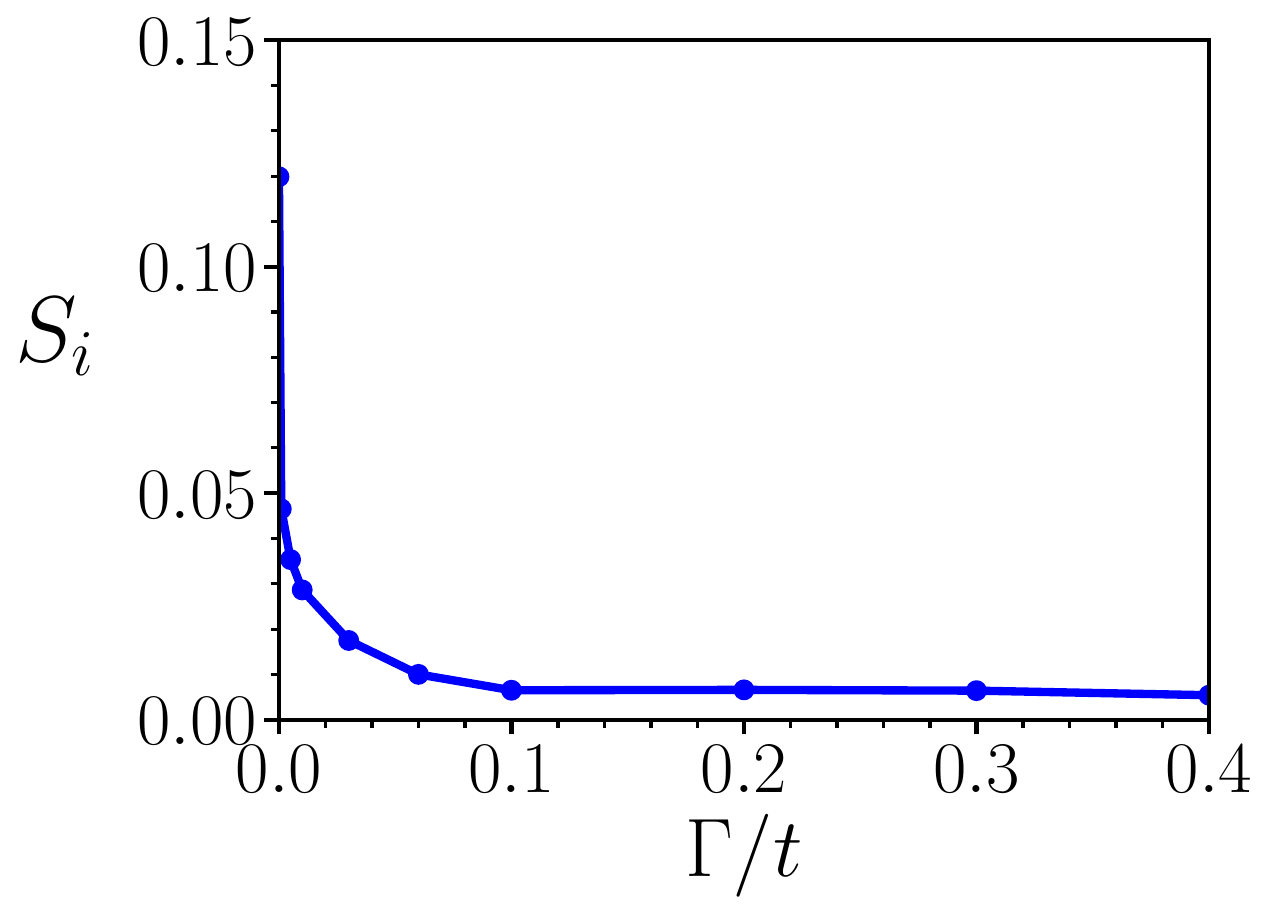}
\caption{The average site spin value of a zigzag edge ($S_i$) is computed. The impurity concentration is 0.1, and the on-site parameter is $U=t$.  Each data point is computed  from $N_D=10$ disorder realizations.
}
\label{spin_TO}
\end{center}	
\end{figure}

Disorder is a singular perturbation\cite{Yang2020} and it significantly impacts the properties of zigzag ribbons, particularly noticeable when measuring the average site spin value along a zigzag edge. Our analysis reveals a dramatic alteration in this value concerning the disorder strength $\Gamma$: a singularity in the slope emerges, as depicted in
Fig. \ref{spin_TO}.

\begin{figure}[hbt!]
\begin{center}
\includegraphics[width = 0.9 \textwidth]{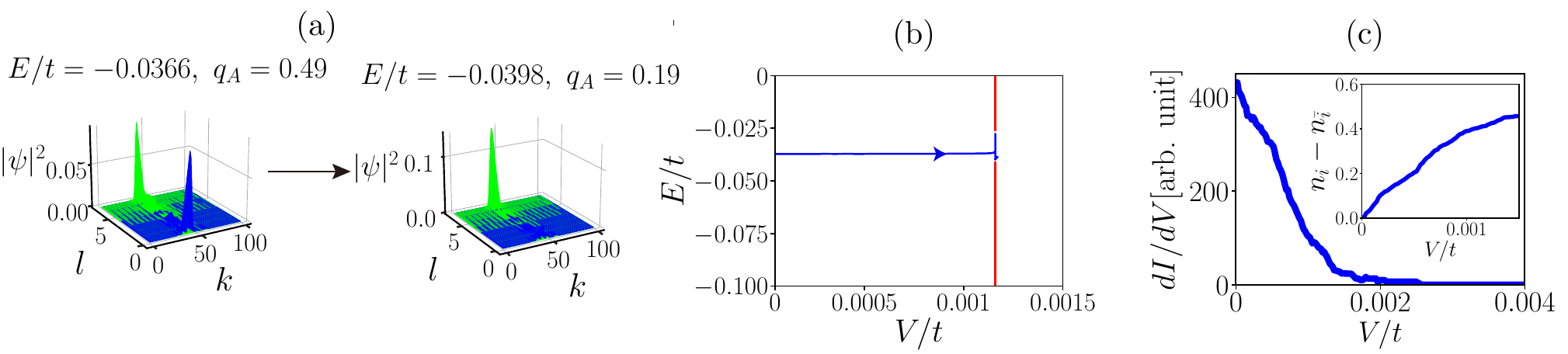}
\caption{(a) Probability densities of fractional and integer charges are plotted before and after the tunneling event, respectively. The two density profiles are generated from a single disorder realization with $V = 0.00115t$ and $V = 0.00116t$. The remaining relevant parameters include $(L, W ) = (100, 8$), and $(U,\Gamma, \delta N ) = (t, 0.01t, 0)$. During this process, we observe the tunneling of approximately one fractional charge to the opposite edge, resulting in the formation of an integer charge. 
(b) The dependence of the energy of a solitonic fractional gap-edge state on $V$ is shown, with the red vertical line indicating specific point where the tunneling event occurs.
 (c) A plot of the tunneling current $I$, proportionate to the total transferred charge (defined as the sum of spin-up and spin-down electrons transferred between edges, $(n_{i} - n_{\bar{i}})$, is displayed as a function of voltage $V$.   
	$n_{i}$ and $n_{\bar{i}}$ represent the average site occupation numbers on opposing zigzag edges, averaged across various disorder realizations and zigzag edge sites. Here, $i$ denotes sites on one zigzag edge, while $\bar{i}$ refers to sites on the opposite zigzag edge. It's important to note that the differential conductance $\frac{dI}{dV}$ at $V=0$ becomes notably large. This plot corresponds to $(U, \Gamma, \delta N) = (t, 0.01t, 0)$ and $(L, W) = (100,8)$.  The disorder configuration is fixed while varying voltage $V$. 
}
\label{Tunneling}
\end{center}	
\end{figure}

Here we also investigate the tunneling of fractional charge between two zigzag edges \cite{Kang, Iyang}. We initiate the investigation by considering a gap-edge state that displays fractional charges in the undoped case, as illustrated in Figure~\ref{Tunneling}(a). We then examine how the probability density of this state evolves as the voltage V between the zigzag edges increases. The transformation of the probability density during the tunneling event is plotted in Figure~\ref{Tunneling}(a). One peak tunnels across the width of the ZGNR and merges with the other peak located at the opposite edge to form an integer charge. Since tunneling of fractional charges occurs within a narrow range of applied voltage $V$, as shown in Figure~\ref{Tunneling}(b), our results suggest the potential existence of a zero-bias anomaly in the differential conductance, see Figure~\ref{Tunneling}(c).

\pagebreak

\section{Crossover phase I}

\begin{figure}[hbt!]
\begin{center}
\includegraphics[width = 0.9 \textwidth]{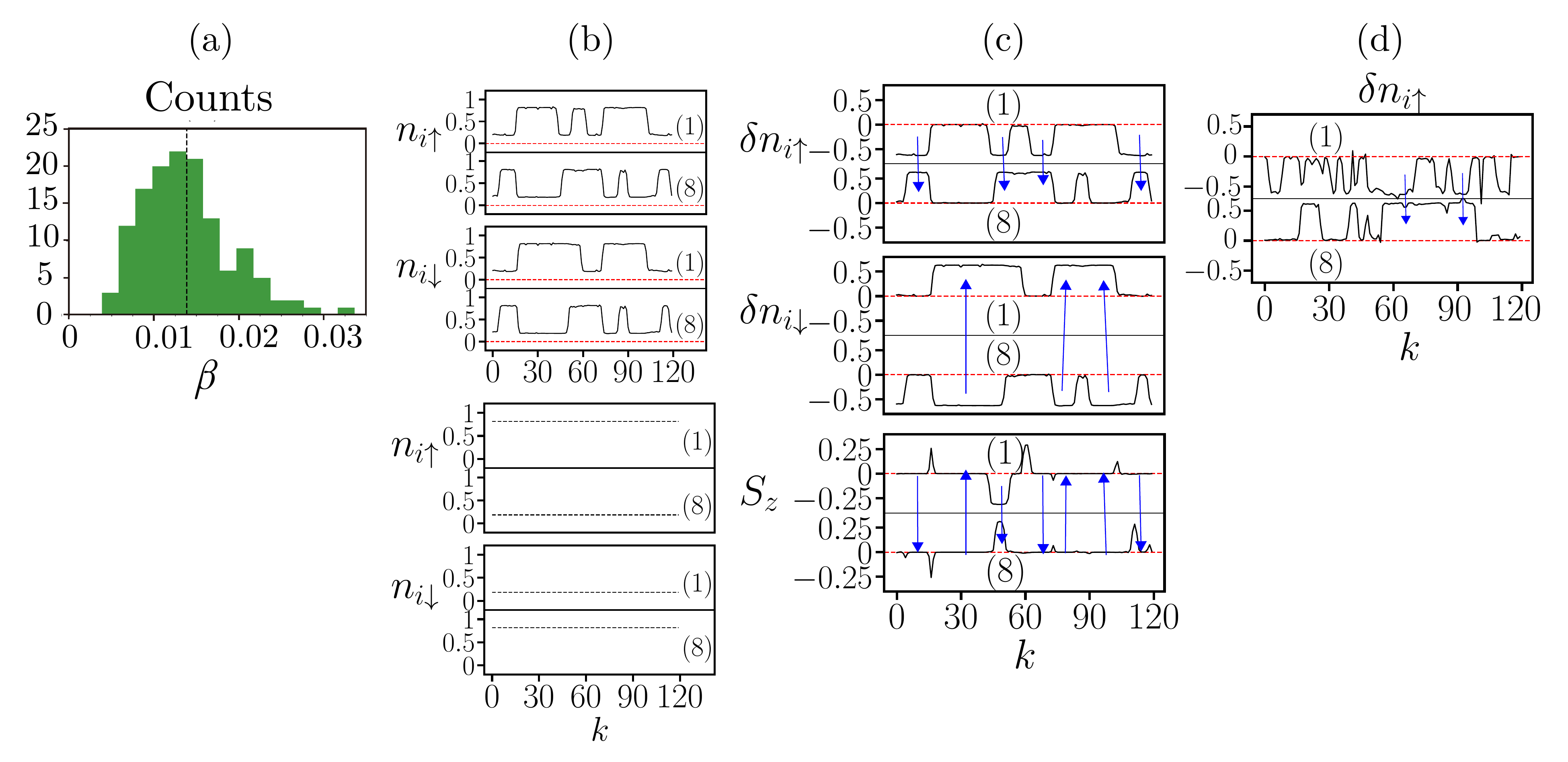}
\caption{(a) Distribution of $\beta$ {for parameters} $ (U,\Gamma, \delta N) = (3t,4.5t,  0)$ with $N_D = 134$. The dashed line indicates the mean value.
(b) $ n_{i,\uparrow (\downarrow)}$ is calculated in a ribbon size $(L, W) = (120, 8)$ with $ (U,\Gamma,  \delta N) = (3t,0.5t,  0)$. The numbers $(1)$ and $(8)$ indicate the upper and lower zigzag edges, as depicted in Figure \ref{TO}(a).  The $x$-axis label $k$ denotes sites along each carbon line. Blue arrows signify locations  where charge transfers occur. (In the lower graph, $ n_{i,\uparrow (\downarrow)}$ are displayed for the non-disorder case in the same condition as the upper graph.)
(c) Change in the spin-up (down) occupation number $\delta n_{i,\uparrow (\downarrow)} = n_{i,\uparrow (\downarrow)}-n^0_{i,\uparrow (\downarrow)}$ of a ribbon size $(L, W) = (120, 8)$ with $ (U,\Gamma,  \delta N,N_D) = (3t,0.5t,  0,1)$. Edge site spins $S_z$ are also shown.  
(d) $\delta n_{i,\uparrow}$ is displayed for a ribbon size $(L, W) = (120,8)$ with $(U,\Gamma,  \delta N,N_D) = (3t, 4t,  0,1)$. In this case, the correlations disappear {as the disorder strength increases.}}
\label{correlation_spin_charge}
\end{center}
\end{figure}

We will provide a comprehensive description of the characteristics of undoped ZGNRs in the crossover I phase, with $U \gtrsim t$ and $U \gtrsim \Gamma$, where the dominant energy is the on-site repulsion $U$. The topologically ordered phase gradually changes into the crossover I phase as $U$ increases, as illustrated in Figure~\ref{phase_diagram_2}(e). In this phase, $\beta$ is nonzero, but its variance is significant, as shown in Figure~\ref{correlation_spin_charge}(a). {The crossover I phase exhibits the following notable properties:} 
(i) The disorder-induced change in the edge occupation numbers $\delta n_{i,\uparrow}= 1/2$ for one type of spin $\sigma$ is {\it entirely} transferred to the opposite edge, i.e., the zigzag  edges are correlated in a nontrivial manner, as shown in Figure~\ref{correlation_spin_charge}(c) for a specific case of $\Gamma/U = 0.17$.    
However, the site positions $k$ on the opposite zigzag edges,    where changes in $\delta n_{i,\uparrow}$ and $\delta n_{i,\downarrow}$  occur do not coincide.
(This is in contrast to the topologically ordered phase where these positions exhibit correlations at nearly identical $k$ values.)
We believe that these edge transfer correlations between zigzag edges modify the ground state entanglement pattern and yield a {\it nonzero fluctuating TEE}. The edge charge transfer  correlations become weaker  when the disorder is stronger (see Figure \ref{correlation_spin_charge}(d) for $\Gamma/t=4$),  leading to a smaller value of $\beta $ (see Figure \ref{phase_diagram_2}(c)).
(ii) Although the zigzag edge changes are fractional, $\delta n_{i,\uparrow}= 1/2$, {as shown in} Figures \ref{correlation_spin_charge}(b) and (c)), the A- and B-probability densities of the mixed chiral gap-edge states responsible for this feature overlap, see  Figure \ref{TO}(b). {Consequently}, fractional charges are ill-defined in this phase.
(iii) The presence of spin-charge separation necessitates that charge transfers for both spin types occur at the same positions $k$. However, these effects are not observed in the crossover I phase.  It is essential to emphasize that the condition $S_z=\frac{1}{2}(n_{i,\uparrow}-n_{i,\downarrow})=0$ at site $i$ is not sufficient for spin-charge separation. To fulfill the conditions, well-defined fractional charges must exist.

\section{Crossover phase II}

\begin{figure}[hbt!]
\begin{center}
\includegraphics[width=0.72\textwidth]{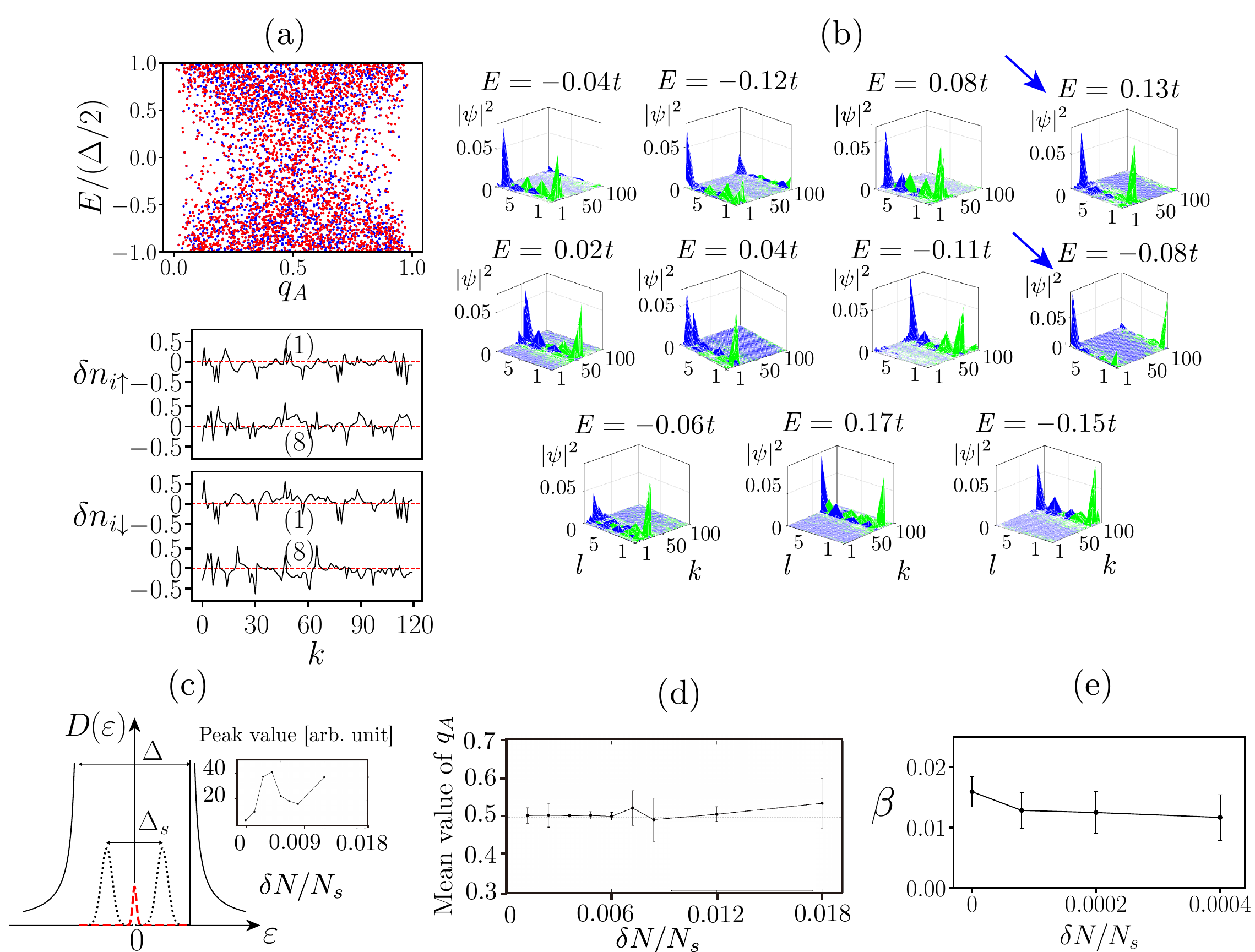}
\caption{(a) (Upper graph) $E$-$ q_A$ diagram in the half-filling case $(U,\Gamma, \delta N) = (t,5t, 0)$, for a ribbon with dimensions $(L, W)  = (100, 8)$, and employing $N_D = 120$ disorder realizations. Red (blue) dots represent states of spin up (down). There are numerous states with $q_A\approx1/2$, but fractional charges barely exist.  
{(Lower graph)} $\delta n_{i,\uparrow(\downarrow)}$ is displayed for a ribbon size $(L,W)=(120,8)$ with $(U,\Gamma, \delta N) = (t,5t,  0)$. (b) Plot of the probability density of several  mixed chiral gap-edge states with $q_A \approx 1/2$ are shown ($N_D = 4$). Green and blue represent densities on two different sublattices. All parameters are similar to those  in (a). Blue arrows point to a gap-edge states that display fractional charges. The 
ribbon has dimensions of $(L, W) = (100, 8)$.
(c)  (Left figure) DOS of an undoped weakly disordered interacting ZGNRs  is shown as a dotted black line, which displays a soft gap~\cite{Yang2019} . The midgap peak of the DOS in the low doping limit~\cite{Yang2021} is marked by the dotted red line. (The Fermi level is above the DOS peak.)  (Right figure) The midgap peak value of the DOS changes as a function of doping concentration $\delta N / N_{\textrm{s}}$, taken  from \cite{Yang2022}). Here, the size of the ribbon  is $(L,W)=(101,8)$ and other parameters are $(\Gamma,U)=(0.01t,t)$.
(d) Mean value of $q_A$ of the midgap states shown in (c) is plotted as a function of doping concentration.
The number of disorder realization is $N_D =200 $. 
(e) Average beta is plotted for several values of doping concentration, {with $(U,\Gamma) = (t, t)$ and 10 disorder realizations for each point}. The analysis includes ribbons of several sizes with dimensions $(L, W, w, l_{\textrm{zig}}, l_{\textrm{arm}})$: $ (170,144, 10, 145, 60)$, $(130,112, 7, 111, 46)$, $(100,64, 4, 81, 28)$, $(80,68, 4, 69, 24)$, and $(50,32, 2, 41, 12)$.}
\label{Fluc1}
\end{center}
\end{figure}

For undoped ZGNRs, another crossover phase emerges when $\Gamma\gg U$ but $U/t\lesssim 1$.  We call this phase crossover II where the disorder strength $\Gamma$ is the dominant energy. As $\Gamma$ increases, the topologically ordered phase undergoes a gradual transition into the crossover phase II, as depicted in Figure \ref{phase_diagram_2}(e). Concurrently, the gap is progressively filled with states, as illustrated in the upper graph of Figure \ref{Fluc1}(a).
Similar to the crossover I phase (Figure \ref{correlation_spin_charge}(a)), $\beta$ is finite with a significant variance. However, in this phase, there are  no charge-transfer correlations between the zigzag edges, as evident from the lower graph in Figure \ref{Fluc1}(a). Some fractional charges {\it may exist}, as shown in Figure \ref{Fluc1}(b), but they are expected to be subject to greater quantum fluctuations.

This observation aligns with the following obtained results:  
(i)   Some  changes  in the edge occupation number are  $\delta n_{i,\sigma} \approx \pm1/2$.
(ii) There are gap-edge  states with  $q_A \approx 1/2$. (Here $
q_A=\sum_{i\in A}|\psi_{i\sigma}(E)|^2$, where $\psi_{i\sigma}(E)$ is the Hartree-Fock eigenstate with energy $E$,  see \cite{Yang2020}.   The probability densities are summed over all sites of the A sublattice.)  However, although the mean value of $q_A$ is $1/2$, its variance in the energy interval $[E-\delta E, E+\delta E]$ is large because $q_A$ varies substantially within this interval, as shown in Figure \ref{Fluc1}(a).
Despite this, a fractional charge of a state in the interval $[E-\delta E,E+\delta E]$  near $E=0$ does {\it not} overlap significantly with the probability densities of other fractional and non-fractional  states in the same energy interval, provided that $\delta E$ is small (for $U\sim t$ and $\Gamma\sim t$, this happens when $\delta E\sim 0.01t$).   (However, in the absence of on-site repulsion, multiple states with the same value of $\delta E$ overlap.) We posit that the interplay between localization and on-site repulsion accounts for these non-overlapping states, as previously discussed at the outset of Sec.\ref{Loc}. Nevertheless, the impact of quantum fluctuations \cite{GV2000} must not be disregarded, as significant overlap with states in the {\it adjacent} energy intervals may occur.  (The Hartree-Fock calculation,  which ignores quantum fluctuations,  may thus overestimate the stability of fractional charges.)


So far, our investigation has primarily focused on the undoped case.   Upon doping, the disorder-free ZGNRs exhibit edge spin density waves, with the opposite edges still being antiferromagnetically coupled. This is in contrast to the edge ferromagnetism observed in undoped ribbons.  If disorder is added to a doped ZGNR,  the spin waves become distorted~\cite{Yang2022}: There is a topological phase transition from modulated ferromagnetic edges at zero doping to distorted spin-wave edges at finite doping. Our results indicated that, under substantial doping, this phase is also a crossover phase II.  
The dependence of the mean value of $q_A$ of the states within the midgap peak  on the number of doped electrons is illustrated in Figure \ref{Fluc1}(d) (the DOS shows a sharp peak at the midgap energy as seen in Figure \ref{Fluc1}(c)). At a low doping concentration, the disorder-averaged value of  $q_A$  of the midgap peak is close to $0.5$, and the midgap states display well-defined fractional charges with small variance, as discussed below Fig.~\ref{correlation_length} (note that the width of the midgap peak is $\delta E\sim 0.005t$).  As the doping concentration increases further, $q_A$ significantly deviates from 0.5, and simultaneously, the DOS midgap peak starts to decrease~\cite{Yang2022}.  These findings imply that even though fractional charges can still be observed, their number decreases with an increase in doping. The gradual change in  $q_A$ as a function of $\delta N / N_\textrm{s}$ indicates that the transition from the phase of the distorted ferromagnetic edge to the phase of the distorted edge spin-wave is not sharp. Figure \ref{Fluc1}(e) shows how $\beta$ decreased with an increase in $\delta N / N_\textrm{s}$.  For a large $\delta N / N_\textrm{s}$, it is computationally demanding to calculate $\beta$ due to the anticipated longer correlation length, as observed in Figure \ref{correlation_length}(c).



\section{Strongly disordered and strongly repulsive phases}

\begin{figure}[hbt!]
\begin{center}
\includegraphics[width=0.77\textwidth]{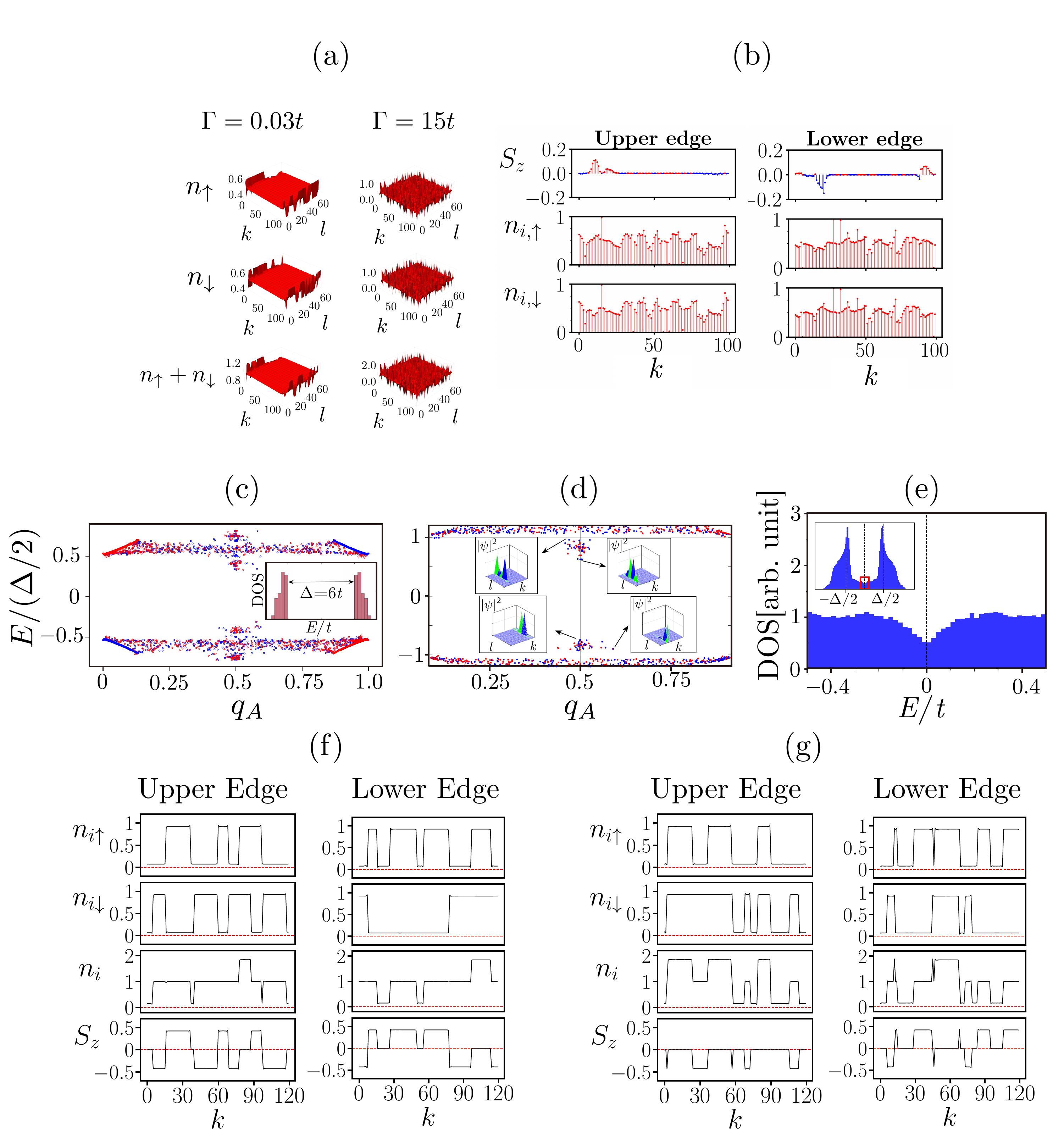}
\caption{(a) Site occupation numbers for spin-up and -down, along with the total occupation numbers, are shown for a ribbon size $(L, W) = (100, 64)$. The calculations were performed for two different sets of parameters: the left one with $(U, \Gamma, \delta N) = (t, 0.03t, 0)$ and the right other with $(U, \Gamma, \delta N) = (t, 15t, 0)$. Both cases were computed by single disorder realization $N_D = 1$.
(b) Edge occupation numbers for spin-up, spin-down, and edge site spins in strongly disordered regime with $(U,\Gamma, \delta N) = (t,15t,0)$. The geometry of the ribbon is $(L,W) = (100,8)$ and these quantities are obtained by single disorder realization $N_D = 1$. 
(c) $E$ - $ q_A$ graph with $(L, W) = (100, 8)$, $ (U,\Gamma,  \delta N) = (7t,0.5t, 0)$ ($N_D = 1$).
Red (blue) dots are states of spin up (down). Solid  red and blue lines  represent  $ q_A$ values of spin-up and -down states  for  $\Gamma = 0$. 
These states correspond to zigzag edge states localized either on the upper or lower zigzag edges. Disorder-free DOS for $ (U,\Gamma,  \delta N) = (7t, 0, 0)$ shown in the inset has a large gap, $\Delta \approx 6t$.
(d) Probability density of several states near gap edges $\pm \Delta/2$ (horizontal thin black lines) with $q_A \approx0.5$ for single disorder realization $N_D = 1$ and $(U,\Gamma, \delta N) = (7t,0.5t,  0)$. {The ribbon's dimensions are set to} $(L,W)=(100,8)$. 
(e) DOS for a ribbon of dimensions $(L, W) = (100, 64)$ with $ (U,\Gamma, \delta N) = (4t, 4t, 0)$ ($N_D = 100$ realizations were made).  The computed TEE value is $\beta = 0$.
(f) Zigzag edge occupation numbers $n_{i,\sigma}$, $n_{i}=n_{i,\uparrow}+n_{i,\downarrow}$ and site spins $S_z$ in the strong repulsive regime with parameters $(U,\Gamma,  \delta N) = (5t,0.01t,  0)$ with a ribbon's dimension of $(L,W)=(120,8)$. These results are based on a single disorder realization ($N_D = 1$). 
(g)  The same quantities are computed for the parameters $(U, \Gamma, \delta N) = (5t, 0.1t, 0)$ using a ribbon with the dimensions mentioned in (f).
At $\Gamma = 0.1t$, more sites exhibit total site occupation numbers $n_{i} \approx 2$ in comparison to $\Gamma = 0.01t$.
}
\label{Fluc2}
\end{center}
\end{figure}

Let us discuss the strongly disordered phase within the region $\Gamma/U\gg 1$ as indicated in Figure \ref{phase_diagram_2}(e). The topological order is destroyed once the disorder strength reaches a sufficiently high level,  for example, $\beta = 0$ at $(U,\Gamma, \delta N) = (t,15t, 0)$. Within this region, both edge charge-transfer correlations  and charge fractionalization are not  well-defined, implying a TEE of zero.
In Figure \ref{Fluc2}(a), site occupation numbers in weak ($\Gamma = 0.03 t$) and strong ($\Gamma = 15 t$) disorder regimes are shown side by side to highlight the difference, where the ones in strong disorder regime highly fluctuate from site to site.
Moreover, edge magnetization is zero almost everywhere (see Figure \ref{Fluc2}(b)). The occupation numbers display sharp values of $n_{i,\sigma}=1$ at some sites, a feature also observed in the density matrix renormalization group calculations of disordered ribbons, as reported in \cite{Yang2022}.

\clearpage

 Another phase emerges in the form of the strongly repulsive phase ($U\gg t$) where fractional charges and zigzag edge correlations become non-existent. As depicted in Figure \ref{phase_diagram_2}(e), $\beta\approx 0$ characterizes this phase.
The $q_A$–$E$ diagram in Figure~\ref{Fluc2}(c) {highlights} the nonperturbative nature of disorder in this regime: the values  $q_A$ are scattered across the entire spectrum between $0$ and $1$ as $\Gamma\rightarrow 0$, whereas they are restricted to the four solid lines at $\Gamma = 0$.  
Also, the $(E, q_A)$ distribution indicates that in this strongly repulsive regime, even with the presence of disorder, a large energy gap persists.
There are no states with $q_A\approx 1/2$ in the vicinity the midgap energy, as illustrated in Figure~\ref{Fluc2}(c). The A- and B-components of the wave function of the states with $q_A\approx 1/2$ near the gap edges $\pm\Delta/2$ overlap,  as revealed in Figure \ref{Fluc2}(d).   
For a stronger disorder (equivalently a larger value of $\Gamma$), the gap becomes increasingly filled with states, leading to a DOS finite at $E = 0$, as shown in Figure \ref{Fluc2}(e).

The main physics characterizing this phase becomes apparent by investigating the zigzag edge structure: the occupation numbers take on values of $n_{i,\sigma} = 1$ or $0$ so charge transfers are one ($\delta n_{i,\sigma}=\pm 1$) in the  strongly repulsive phase (the total site occupation number of each site is $n_{i}\approx 2$, $1$, or $0$ despite a strong on-site repulsion $U$). It is important to note that no transfer of fractional charges was observed between zigzag edges.  This absence of charge transfer is attributed to the absence of mixed chiral gap-edge states. Additionally, it is worth mentioning that the edge magnetization displays sharp domain walls, as evidenced in Figures \ref{Fluc2}(f)-(g).

\section{Summary and Discussion}

We computed the phase diagram of zigzag graphene nanoribbons as a function of the on-site repulsion $U$, doping $\delta N$,  and disorder strength $\Gamma$.
Our analysis identified the universal, crossover, strongly disordered, and strongly repulsive phases.
Each phase of the phase diagram  was defined by the TEE value and its variance.  We also  investigated how the values of the TEE are related to  the following physical properties:  the presence of charge fractionalization and  the edge  charge transfer correlations   between the opposite zigzag edges.  When both of these properties are present, along with correlations leading to spin-charge separation, $\beta$ was universal. If only one of these properties was present, $\beta$ was nonuniversal and its variance  was significant. However, when  both properties were entirely absent, $\beta$ was approximately zero. These  results suggest that non-local correlations in zigzag ribbons give rise to non-zero TEE values. Furthermore, our investigation identified a  strongly repulsive  phase with zero TEE in large on-site repulsion and weak disorder limits.  Its ground state contains abrupt kinks in zigzag edge magnetizations without charge fractionalization, which is a consequence of the singular perturbative nature of  the disorder potential.
An additional phase with zero TEE was identified, corresponding to the strongly disordered regime $\Gamma\gg U$. In this phase, the edge site occupation numbers fluctuate highly from site to site, and antiferromagnetic coupling between the two edges is nearly destroyed. Each phase of the phase diagram presents a distinct zigzag-edge structure.

We also conducted an investigation into the effect of the interplay between localization and on-site repulsion on the charge quantization. In low-doped and/or weakly disordered  ZGNRs, this  interplay  contributes to the spatial separation of quasi-degenerate gap-edge states, and protects the charge fractionalization against quantum fluctuations. Even in the presence of moderately strong disorder, charge fractionalization is not completely eradicated.

 Reference~\cite{Chen2017} showed that, in the absence of disorder, there can be several different antiferromagnetic and ferromagnetic phase changes as width and doping concentration increase.  In our work, we have  investigated ribbons with widths smaller  than $\sim 10$ nm and doping concentrations less than $\sim 0.003$. We focused on the low doping region of disordered ribbons because our investigation shows that topological order is stable only in that regime.  In this low doping region,   each edge develops a spin density wave~\cite{Yang2022}. However, the \textit{average} site spin values  of the opposite edges are still being antiferromagnetically coupled.


We briefly discuss some experimental implications.  Investigating the tunneling of fractional charges between the zigzag edges of undoped, topologically ordered zigzag ribbons under an applied voltage could be an interesting avenue, similar to experiments performed on fractional quantum Hall edges~\cite{Kang, Iyang}.
 Such an effect may lead to a zero-bias anomaly. Additionally, observing the presence of nonlocal charge transfers between the zigzag edges of the crossover phase I would be of interest. This can be investigated by measuring correlations between the edge site occupation numbers using a scanning tunneling microscope~\cite{Andrei}. In the crossover phase II, fractional $e^-/2$ edge charges are present; however, unusual transport and magnetic susceptibility properties are not expected due to the absence of spin-charge separation. (In contrast, the topologically ordered phase is expected to display unusual transport and magnetic susceptibility because of spin-charge separation~\cite{Chung,Yang2020}.) Similarly, using scanning tunneling microscopy can help to validate the predicted edge occupation numbers in the strongly repulsive and strongly disordered phases.


\newpage
\section*{References}
\begin{thebibliography}{10}

\bibitem{Lei13}
Jon~Magne Leinaas and Jan Myrheim.
\newblock ``On the theory of identical particles''.
\newblock \href{https://dx.doi.org/10.1007/BF02727953}{Nuovo Cimento B {\bf
  37}, 1--23}~(1977).

\bibitem{Wilczek03}
Frank Wilczek.
\newblock ``Quantum mechanics of fractional-spin particles''.
\newblock \href{https://dx.doi.org/10.1103/PhysRevLett.49.957}{Phys. Rev. Lett.
  {\bf 49}, 957--959}~(1982).

\bibitem{Arovas}
Daniel Arovas, J.~R. Schrieffer, and Frank Wilczek.
\newblock ``Fractional statistics and the quantum hall effect''.
\newblock \href{https://dx.doi.org/10.1103/PhysRevLett.53.722}{Phys. Rev. Lett.
  {\bf 53}, 722--723}~(1984).

\bibitem{Nakamura01}
J.~Nakamura, S.~Liang, G.~C. Gardner, and M.~J. Manfra.
\newblock ``Direct observation of anyonic braiding statistics''.
\newblock
  \href{https://dx.doi.org/https://doi.org/10.1038/s41567-020-1019-1}{Nature
  Physics {\bf 16}, 931--936}~(2020).

\bibitem{Barto1}
H.~Bartolomei, M.~Kumar, R.~Bisognin, A.~Marguerite, J.-M. Berroir,
  E.~Bocquillon, B.~Pla{\c c}ais, A.~Cavanna, Q.~Dong, U.~Gennser, Y.~Jin, and
  G.~F{\`e}ve.
\newblock ``Fractional statistics in anyon collisions''.
\newblock \href{https://dx.doi.org/10.1126/science.aaz5601}{Science {\bf 368},
  173--177}~(2020).

\bibitem{GV2019}
Steven~M Girvin and Kun Yang.
\newblock ``Modern condensed matter physics''.
\newblock \href{https://dx.doi.org/10.1017/9781316480649}{Cambridge University
  Press}. Cambridge~(2019).

\bibitem{Pach}
Jianis~K Pachos.
\newblock ``Introduction to topological quantum computation''.
\newblock \href{https://dx.doi.org/10.1017/CBO9780511792908}{Cambridge
  University Press}. Cambridge~(2012).

\bibitem{Wen11}
Xiao-Gang Wen.
\newblock ``Colloquium: Zoo of quantum-topological phases of matter''.
\newblock \href{https://dx.doi.org/10.1103/RevModPhys.89.041004}{Rev. Mod.
  Phys. {\bf 89}, 041004}~(2017).

\bibitem{NPlaugh}
R.~B. Laughlin.
\newblock ``Anomalous quantum hall effect: An incompressible quantum fluid with
  fractionally charged excitations''.
\newblock \href{https://dx.doi.org/10.1103/PhysRevLett.50.1395}{Phys. Rev.
  Lett. {\bf 50}, 1395--1398}~(1983).

\bibitem{Yang}
S.-R.~Eric Yang.
\newblock ``Topologically ordered zigzag nanoribbon''.
\newblock \href{https://dx.doi.org/10.1142/13013}{World Scientific, Singapore}.
  ~(2023).

\bibitem{Kitaev11}
Alexei Kitaev and John Preskill.
\newblock ``Topological entanglement entropy''.
\newblock \href{https://dx.doi.org/10.1103/PhysRevLett.96.110404}{Phys. Rev.
  Lett. {\bf 96}, 110404}~(2006).

\bibitem{Levin11}
Michael Levin and Xiao-Gang Wen.
\newblock ``Detecting topological order in a ground state wave function''.
\newblock \href{https://dx.doi.org/10.1103/PhysRevLett.96.110405}{Phys. Rev.
  Lett. {\bf 96}, 110405}~(2006).

\bibitem{Haldane191}
Hui Li and F.~D.~M. Haldane.
\newblock ``Entanglement spectrum as a generalization of entanglement entropy:
  Identification of topological order in non-abelian fractional quantum hall
  effect states''.
\newblock \href{https://dx.doi.org/10.1103/PhysRevLett.101.010504}{Phys. Rev.
  Lett. {\bf 101}, 010504}~(2008).

\bibitem{Altshuler}
B.~Altshuler.
\newblock ``Inductory anderson localization''.
\newblock In Advanced Workshop on Anderson Localization, Nonlinearity and
  Turbulence: a Cross-Fertilization.
\newblock International Centre for Theoretical Physics~(2010).

\bibitem{GV2000}
Steven~M Girvin.
\newblock ``The quantum hall effect: novel excitations and broken symmetries''.
\newblock In Aspects topologiques de la physique en basse dimension.
  Topological aspects of low dimensional systems.
\newblock Pages 53--175.
\newblock Springer~(1999).

\bibitem{Fujita}
Mitsutaka Fujita, Katsunori Wakabayashi, Kyoko Nakada, and Koichi Kusakabe.
\newblock ``Peculiar localized state at zigzag graphite edge''.
\newblock \href{https://dx.doi.org/10.1143/JPSJ.65.1920}{J. Phys. Soc. Jpn.
  {\bf 65}, 1920--1923}~(1996).

\bibitem{Brey2006}
L.~Brey and H.~A. Fertig.
\newblock ``Electronic states of graphene nanoribbons studied with the dirac
  equation''.
\newblock \href{https://dx.doi.org/10.1103/PhysRevB.73.235411}{Phys. Rev. B
  {\bf 73}, 235411}~(2006).

\bibitem{Lyang}
Li~Yang, Cheol-Hwan Park, Young-Woo Son, Marvin~L. Cohen, and Steven~G. Louie.
\newblock ``Quasiparticle energies and band gaps in graphene nanoribbons''.
\newblock \href{https://dx.doi.org/10.1103/PhysRevLett.99.186801}{Phys. Rev.
  Lett. {\bf 99}, 186801}~(2007).

\bibitem{Pisa1}
L.~Pisani, J.~A. Chan, B.~Montanari, and N.~M. Harrison.
\newblock ``Electronic structure and magnetic properties of graphitic
  ribbons''.
\newblock \href{https://dx.doi.org/10.1103/PhysRevB.75.064418}{Phys. Rev. B
  {\bf 75}, 064418}~(2007).

\bibitem{Cai2}
Pascal Ruffieux, Shiyong Wang, Bo~Yang, Carlos S{\'a}nchez-S{\'a}nchez, Jia
  Liu, Thomas Dienel, Leopold Talirz, Prashant Shinde, Carlo~A Pignedoli,
  Daniele Passerone, et~al.
\newblock ``On-surface synthesis of graphene nanoribbons with zigzag edge
  topology''.
\newblock \href{https://dx.doi.org/10.1038/nature17151}{Nature {\bf 531},
  489--492}~(2016).

\bibitem{Kolmer}
Marek Kolmer, Ann-Kristin Steiner, Irena Izydorczyk, Wonhee Ko, Mads Engelund,
  Marek Szymonski, An-Ping Li, and Konstantin Amsharov.
\newblock ``Rational synthesis of atomically precise graphene nanoribbons
  directly on metal oxide surfaces''.
\newblock \href{https://dx.doi.org/10.1126/science.abb8880}{Science {\bf 369},
  571--575}~(2020).

\bibitem{Brey}
Luis Brey, Pierre Seneor, and Antonio Tejeda, editors.
\newblock ``Graphene nanoribbons''.
\newblock \href{https://dx.doi.org/10.1088/978-0-7503-1701-6}{2053-2563}. IOP
  Publishing. ~(2019).

\bibitem{Heeger}
A.~J. Heeger, S.~Kivelson, J.~R. Schrieffer, and W.~P. Su.
\newblock ``Solitons in conducting polymers''.
\newblock \href{https://dx.doi.org/10.1103/RevModPhys.60.781}{Rev. Mod. Phys.
  {\bf 60}, 781--850}~(1988).

\bibitem{Yang2019}
Y.~H. Jeong, \text{S.-R. Eric Yang}, and Min-Chul Cha.
\newblock ``Soliton fractional charge of disordered graphene nanoribbon''.
\newblock \href{https://dx.doi.org/10.1088/1361-648X/ab146b}{Journal of
  Physics: Condensed Matter {\bf 31}, 265601}~(2019).

\bibitem{yang1}
\text{S.-R. Eric Yang}.
\newblock ``Soliton fractional charges in graphene nanoribbon and
  polyacetylene: similarities and differences''.
\newblock \href{https://dx.doi.org/10.3390/nano9060885}{Nanomaterials {\bf 9},
  885}~(2019).

\bibitem{Yang2020}
S.-R.~Eric Yang, Min-Chul Cha, Hye~Jeong Lee, and Young~Heon Kim.
\newblock ``Topologically ordered zigzag nanoribbon: $e/2$ fractional edge
  charge, spin-charge separation, and ground-state degeneracy''.
\newblock \href{https://dx.doi.org/10.1103/PhysRevResearch.2.033109}{Phys. Rev.
  Research {\bf 2}, 033109}~(2020).

\bibitem{Dob}
Vladimir Dobrosavljevic, Nandini Trivedi, and James~M. Valles, Jr.
\newblock ``{Conductor-Insulator Quantum Phase Transitions}''.
\newblock
  \href{https://dx.doi.org/10.1093/acprof:oso/9780199592593.001.0001}{Oxford
  University Press}. ~(2012).

\bibitem{Belitz}
D.~Belitz and T.~R. Kirkpatrick.
\newblock ``The anderson-mott transition''.
\newblock \href{https://dx.doi.org/10.1103/RevModPhys.66.261}{Rev. Mod. Phys.
  {\bf 66}, 261--380}~(1994).

\bibitem{Byczuk}
Krzysztof Byczuk, Walter Hofstetter, and Dieter Vollhardt.
\newblock ``Competition between anderson localization and antiferromagnetism in
  correlated lattice fermion systems with disorder''.
\newblock \href{https://dx.doi.org/10.1103/PhysRevLett.102.146403}{Phys. Rev.
  Lett. {\bf 102}, 146403}~(2009).

\bibitem{Yang2021}
Young~Heon Kim, Hye~Jeong Lee, and \text{S.-R. Eric Yang}.
\newblock ``Topological entanglement entropy of interacting disordered zigzag
  graphene ribbons''.
\newblock \href{https://dx.doi.org/10.1103/PhysRevB.103.115151}{Phys. Rev. B
  {\bf 103}, 115151}~(2021).

\bibitem{Yang2022}
Young~Heon Kim, Hye~Jeong Lee, Hyun-Yong Lee, and \text{S.-R. Eric Yang}.
\newblock ``New disordered anyon phase of doped graphene zigzag nanoribbon''.
\newblock \href{https://dx.doi.org/10.1038/s41598-022-18731-6}{Scientific
  Reports {\bf 12}, 14551}~(2022).

\bibitem{Efros}
Al~L {\'E}fros and Boris~I Shklovskii.
\newblock ``Coulomb gap and low temperature conductivity of disordered
  systems''.
\newblock \href{https://dx.doi.org/10.1088/0022-3719/8/4/003}{Journal of
  Physics C: Solid State Physics {\bf 8}, L49--L51}~(1975).

\bibitem{Lima2012}
Leandro R.~F. Lima, Felipe~A. Pinheiro, Rodrigo~B. Capaz, Caio~H. Lewenkopf,
  and Eduardo~R. Mucciolo.
\newblock ``Effects of disorder range and electronic energy on the perfect
  transmission in graphene nanoribbons''.
\newblock \href{https://dx.doi.org/10.1103/PhysRevB.86.205111}{Phys. Rev. B
  {\bf 86}, 205111}~(2012).

\bibitem{Canri}
G.~S. Canright, S.~M. Girvin, and A.~Brass.
\newblock ``Superconductive pairing of fermions and semions in two
  dimensions''.
\newblock \href{https://dx.doi.org/10.1103/PhysRevLett.63.2295}{Phys. Rev.
  Lett. {\bf 63}, 2295--2298}~(1989).

\bibitem{Huang2011}
Ching-Yu Huang and Feng-Li Lin.
\newblock ``Topological order and degenerate singular value spectrum in
  two-dimensional dimerized quantum heisenberg model''.
\newblock \href{https://dx.doi.org/10.1103/PhysRevB.84.125110}{Phys. Rev. B
  {\bf 84}, 125110}~(2011).

\bibitem{Stau}
T.~Stauber, P.~Parida, M.~Trushin, M.~V. Ulybyshev, D.~L. Boyda, and
  J.~Schliemann.
\newblock ``Interacting electrons in graphene: Fermi velocity renormalization
  and optical response''.
\newblock \href{https://dx.doi.org/10.1103/PhysRevLett.118.266801}{Phys. Rev.
  Lett. {\bf 118}, 266801}~(2017).

\bibitem{Neto}
A.~H. Castro~Neto, F.~Guinea, N.~M.~R. Peres, K.~S. Novoselov, and A.~K. Geim.
\newblock ``The electronic properties of graphene''.
\newblock \href{https://dx.doi.org/10.1103/RevModPhys.81.109}{Rev. Mod. Phys.
  {\bf 81}, 109--162}~(2009).

\bibitem{Yang1995}
\text{S.-R. Eric Yang}, A.~H. MacDonald, and Bodo Huckestein.
\newblock ``Interactions, localization, and the integer quantum hall effect''.
\newblock \href{https://dx.doi.org/10.1103/PhysRevLett.74.3229}{Phys. Rev.
  Lett. {\bf 74}, 3229--3232}~(1995).

\bibitem{Mac1}
\text{S.-R. Eric Yang} and A.~H. MacDonald.
\newblock ``Coulomb gaps in a strong magnetic field''.
\newblock \href{https://dx.doi.org/10.1103/PhysRevLett.70.4110}{Phys. Rev.
  Lett. {\bf 70}, 4110--4113}~(1993).

\bibitem{Peschel119}
Ingo Peschel.
\newblock ``Calculation of reduced density matrices from correlation
  functions''.
\newblock \href{https://dx.doi.org/10.1088/0305-4470/36/14/101}{Journal of
  Physics A: Mathematical and General {\bf 36}, L205}~(2003).

\bibitem{Bal}
H-C. Jiang, Z.~Wang, and L.~Balents.
\newblock ``Identifying topological order by entanglement entropy''.
\newblock \href{https://dx.doi.org/10.1038/nphys2465}{Nature Phys {\bf 8},
  902--905}~(2012).
  
\bibitem{Chen2017}
W.-C. Chen, Y. Zhou, S.-L. Yu, W.-G. Yin, and C.-D. Gong.
\newblock ``Width-Tuned Magnetic Order Oscillation on Zigzag Edges of Honeycomb Nanoribbons''.
\newblock \href{https://dx.doi.org/10.1021/acs.nanolett.7b01474}{Nano Lett., {\bf 17}(7),
	4400--4404}~(2017).



\bibitem{Kang}
W~Kang, HL~Stormer, LN~Pfeiffer, KW~Baldwin, and KW~West.
\newblock ``Tunnelling between the edges of two lateral quantum hall systems''.
\newblock \href{https://dx.doi.org/https://doi.org/10.1038/47436}{Nature {\bf
  403}, 59--61}~(2000).

\bibitem{Iyang}
I.~Yang, W.~Kang, K.~W. Baldwin, L.~N. Pfeiffer, and K.~W. West.
\newblock ``Cascade of quantum phase transitions in tunnel-coupled edge
  states''.
\newblock \href{https://dx.doi.org/10.1103/PhysRevLett.92.056802}{Phys. Rev.
  Lett. {\bf 92}, 056802}~(2004).

\bibitem{Andrei}
Eva~Y Andrei, Guohong Li, and Xu~Du.
\newblock ``Electronic properties of graphene: a perspective from scanning
  tunneling microscopy and magnetotransport''.
\newblock \href{https://dx.doi.org/10.1088/0034-4885/75/5/056501}{Reports on
  Progress in Physics {\bf 75}, 056501}~(2012).

\bibitem{Chung}
T.-C. Chung, F.~Moraes, J.~D. Flood, and A.~J. Heeger.
\newblock ``Solitons at high density in trans-${(\mathrm{CH})}_{x}$: Collective
  transport by mobile, spinless charged solitons''.
\newblock \href{https://dx.doi.org/10.1103/PhysRevB.29.2341}{Phys. Rev. B {\bf
  29}, 2341--2343}~(1984).

\end{thebibliography}

\section*{Acknowledgements}
S.R.E.Y. was supported by the Basic Science Research Program of the National Research Foundation of
Korea (NRF), funded by the Ministry of Science and ICT (MSIT) NRF-2021R1F1A1047759.   Two other grants are also acknowledged:  BK21 FOUR (Fostering Outstanding Universities for Research)
and KISTI Supercomputing Center with supercomputing resources including technical support KSC-2022-CRE-0345.
%
%
%
%
%
%
%


\end{document}